\documentclass[a4paper,twoside]{article}

\usepackage{epsfig}
\usepackage{subfigure}
\usepackage{calc}
\usepackage{amssymb}
\usepackage{amstext}
\usepackage{amsmath}
\usepackage{amsthm}
\usepackage{multicol}
\usepackage{pslatex}
\usepackage{apalike}
\usepackage{epsfig,url,graphicx,tabulary}
\usepackage{tikz}
\usepackage{SCITEPRESS} 
\usepackage[small]{caption}

\definecolor{specialblue}{RGB}{0,124,232}
\begin{document}

\title{Densifying the sparse cloud SimSaaS: The need of a synergy among agent-directed simulation, SimSaaS and HLA}


\author{\authorname{Tiago Azevedo, Rosaldo J. F. Rossetti, Jorge G. Barbosa}
\affiliation{Artificial Intelligence and Computer Science Lab\\
Department of Informatics Engineering\\
Faculty of Engineering, University of Porto, Portugal}
\email{\{tiago.manuel, rossetti, jbarbosa\}@fe.up.pt}
}

\keywords{Agent-directed simulation, Agent-supported simulation, HLA, High Level Architecture, Cloud, SimSaaS, Simulation Software-as-a-service}

\abstract{Modelling \& Simulation (M\&S) is broadly used in real scenarios where making physical modifications could be highly expensive. With the so-called Simulation Software-as-a-Service (SimSaaS), researchers could take advantage of the huge amount of resource that cloud computing provides. Even so, studying and analysing a problem through simulation may need several simulation tools, hence raising interoperability issues. Having this in mind, IEEE developed a standard for interoperability among simulators named High Level Architecture (HLA). Moreover, the multi-agent system approach has become recognised as a convenient approach for modelling and simulating complex systems. Despite all the recent works and acceptance of these technologies, there is still a great lack of work regarding synergies among them. This paper shows by means of a literature review this lack of work or, in other words, the sparse Cloud SimSaaS. The literature review and the resulting taxonomy  are the main contributions of this paper, as they provide a research agenda illustrating future research opportunities and trends.}

\onecolumn \maketitle \normalsize \vfill

\section{\uppercase{Introduction}}

\noindent Modelling \& Simulation (M\&S) is widely used in real scenarios such as traffic and transportation networks, where making physical modifications could be highly expensive, dependent on political decisions and very disruptive to the environment. Its uses could be decision making and what-if analysis, performance optimisations, testing and training, making M\&S methodologies a huge need for Universities and companies worldwide.

Nowadays, there is a new paradigm called Simulation Software-as-a-Service (SimSaaS) where simulation software is used in the form of services, thanks to the latest evolutions in cloud computing and Software-as-a-Service (SaaS). Instead of having the simulation software installed on their own computers, researchers could take advantage of the huge amount of resource that cloud computing provides.

Even so, studying and analysing a problem through simulation may need several simulation tools, with different resolutions and domain perspectives, hence raising interoperability issues that could not be trivial to solve.

IEEE has already a standard for interoperability among simulators named High Level Architecture (HLA). HLA is covered by many works in the literature. Moreover, the multi-agent system approach has become recognised as a convenient approach for modelling and simulating complex systems~\cite{moya2007towards}. Indeed, many researchers have developed work regarding agents.

This paper shows that the M\&S community did not make a complete jump from the simulations in the local machines to the simulations in the cloud offered in the form of services. To prove such an assertion, we conducted a literature review to make the body of knowledge of the current synergy among agent-directed simulation, SimSaaS and HLA. 

The literature review was conducted using the methodological and systematic framework proposed by~\cite{VomBrockeSNRPC2009Reconstructing}. It is not important in the context of this paper to explicitly describe all the phases. Yet, some considerations must be made. The databases sources selected were Scopus, Engineering Village and ACM. These databases are commonly known to contain vast work and have been used by many researchers in software engineering. The queries made to the databases sources were based in four main keywords: SimSaaS, Cloud Computing, HLA and agents. It is considered a time frame from 2004 to 2015.  The evolution of knowledge and technology in the software engineering field is tremendous every year. Thus, a time frame of a decade seems enough. 

This paper will start to briefly explain some preliminary background concepts regarding the agent-oriented paradigm, HLA and cloud for a better understanding of the scope of this work. After that, the results of the literature review are broadly indicated and a taxonomy of the research work is presented. The literature review and resulting taxonomy are the main contributions of this paper, as they provide a research agenda illustrating future research opportunities and trends. 

\section{\uppercase{Preliminary Background}}

For the sake of clarification in future references, this section will briefly describe what is the agent-directed simulation paradigm, the HLA standard and the cloud paradigm.

Yilmaz and \"{O}ren~\cite{yilmaz2007agent} indicated that the agent-directed simulation paradigm consists in three main areas: (1)~simulation for agents (simulation of agent systems, that is, the simulation model is one or more agents), (2)~agent-based simulation (model behaviour generation or monitoring of this process by using agents) and (3)~agent-supported simulation (improving simulation by using agents as support facilities). There are several researchers which consider \textit{agent simulation} and \textit{agent-based simulation} the same principle as they do not take into account the contribution of agents in model generation. In this work, it is adopted the same perspective in which the two principles are seen as the same.

In order to have a structural basis for interoperability among simulators, IEEE developed HLA, a software standard that provides a common technical architecture for distributed M\&S. A federate is the name given to every participant of the simulation, whereas each one can interact within a federation. Communication between simulators is possible thanks to a Run-Time Infrastructure (RTI). HLA baseline components are: (1)~Federate Interface Specification~\cite{ieee2010hlaFederate}; (2)~Framework and Rules~\cite{ieee2010hlaRUules}; and (3)~Object Model Template (OMT) Specification~\cite{ieee2010hlaOMT}. The first is a definition of the services that each federate can use for communication. The second is a set of rules that ensure the proper interaction within a federation. The latter is a specification of the format and syntax of the data that is exchanged among federates.

Cloud computing is a fresh and on-going recent buzzword where more and more work is being done not only in the industry but also among academics. Nonetheless, there is no general consensus on an unambiguous definition~\cite{geelan2009twenty}. A problem in defining cloud computing is that it overlaps with other domains in distributed systems. Foster et al.~\cite{foster2008cloud} define the fields of distributed systems according to scale and domain (application-oriented versus service-oriented). Web 2.0 covers the spectrum of service-oriented applications, opposing to the Supercomputing and Cluster Computing, which have been more focused on traditional local applications. Cloud Computing lies at the large-scale side, being more scalable than Grid Computing. Grid Computing overlaps with all these fields, and because of that it is normal to exist wrong definitions.

Despite the overlapping of cloud computing with other domains, it is possible to distinguish it from grid computing. Indeed, there are three aspects that are new in cloud computing~\cite{armbrust2010view}: (1)~the appearance of infinite computing resources available on demand; (2)~the elimination of an up-front commitment by cloud users; and (3)~the ability to pay for use of computing resources on a short-term basis as needed. Two new buzzwords emerged from Cloud Computing, trying to extend it even further: Fog Computing~\cite{bonomi2012fog} and Cloud 2.0~\cite{miluzzo2014m}. Concluding, we adopt the definition provided by The National Institute of Standards and Technology~\cite{mell2011nist}: \textit{Cloud computing is a model for enabling ubiquitous, convenient, on-demand network access to a shared pool of configurable computing resources (e.g., networks, servers, storage, applications, and services) that can be rapidly provisioned and released with minimal management effort or service provider interaction.}

\section{\uppercase{An Unexplored Cloud SimSaaS}}

This section will show the unexplored Cloud SimSaaS by means of a literature review. It starts by referring SimSaaS and Cloud generically, following the HLA standard and the agent-directed simulation topics.

In our research, it was clear that SimSaaS is a very modern topic. The great majority of the papers were published after 2011 and the first ones do not directly use the term \textit{SimSaaS}, vaguely mentioning \textit{simulation} and \textit{web}. Although there is an increasing number of works per year, papers about SimSaaS are not many. Nevertheless, they are wide concerning the domains of application. It is possible to see works in the biomedical domain~\cite{Sawicki20121190}, in crowd and pedestrian field~\cite{Wang:2015:SSM:2723553.2723554} as well as works regarding ontology learning~\cite{Wang2014179}, traffic and transportation~\cite{Harri2010}, scheduling parallel discrete event simulation jobs~\cite{Liu2012} and a cloud simulation in manufacturing~\cite{Taylor201489}, just to cite some.

Beyond these specific domain works, there are also some generic ones concerning frameworks for development, for example~\cite{Tsai:2011:SSS:2048370.2048381}~\cite{Guo2011113}. Moreover, Cayirci refers to SimSaaS using the term \textit{Modelling and Simulation-as-a-service} (MSaaS). He clarifies MSaaS, including top threats~\cite{Cayirci2013389}. He also talks about the notions and relations of accountability, risk and trust modelling~\cite{Cayirci20131347}, as well as MSaaS composition in multi-datacenter or multi-cloud scenarios~\cite{Cayirci20131388}. Cayirci is not the only one using the term MSaaS:~\cite{Siegfried2014248} illustrate potential benefits that may be achieved by MSaaS and challenges that remain to be solved.

\begin{figure}[h!]
\centering
\includegraphics[width=.9\linewidth]{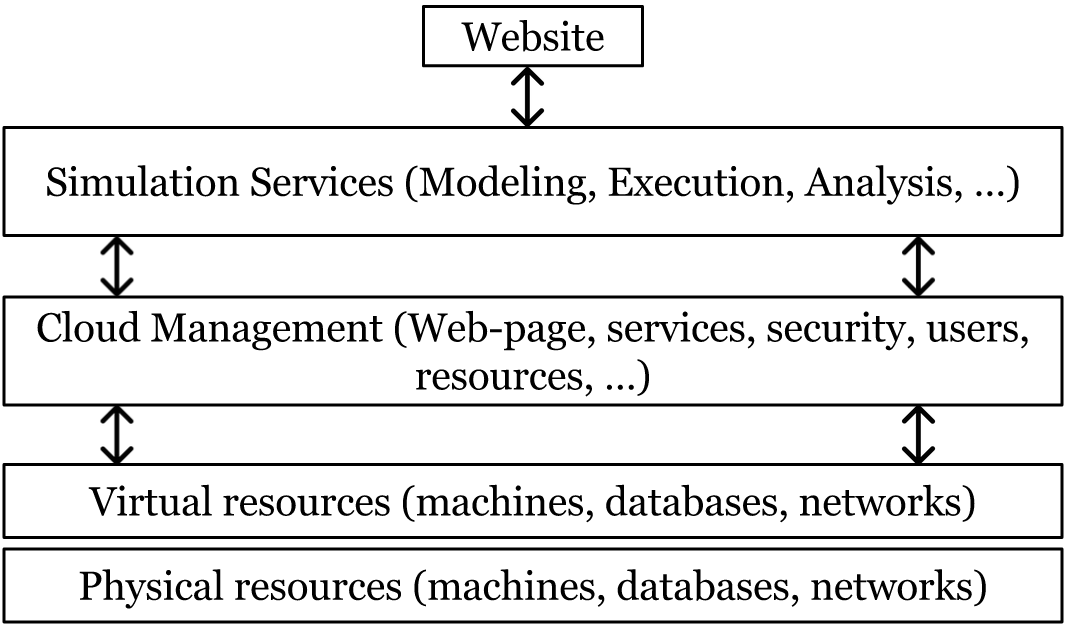}
\caption{ The main blocks of the Cloud Simulation general architecture~\protect\cite{Liu201271}}
\label{fig:csim_arch}
\vspace{-0.9em}
\end{figure}

Since cloud simulation started to be studied, an overall picture took some time to arrive. So, in 2012 Liu et al.~\cite{Liu201271} proposed a general architecture of cloud simulation in the form of a SaaS type cloud. Figure~\ref{fig:csim_arch} summarises the main blocks of this architecture. The simulation services which are offered through a website to very different users, are divided into three self-explanatory groups: Modelling as a Service, Execution as a Service and Analysis as a Service. All these services are possible due to the physical and virtual resources in the bottom, as well as the so-called Cloud Operating System which manages and connects the baseline infrastructure to the top.

Although HLA has already been used in a variety of works like agent-based simulations, there is almost no clear references about interoperability among simulators when talking about SimSaaS. A first approach in extending HLA to support grid-wide distributed simulation dates back to 2005~\cite{Xie:2005:SPH:1069810.1070184}. 

A very relevant work from 2012 discusses how HLA can be integrated with Service Oriented Architecture (SOA) in the context of a smart building project~\cite{Dragoicea:2012:IHS:2310096.2310200}. The Simulation Engine Service is exposed by RESTful services. Inside this Engine, there is also a RESTful API that exposes access to the RTI's federation management and deals with the creation, initialisation, deletion, starting, stopping, and execution of simulations. Still in this context, the authors refer another paper~\cite{Wang:2008:SHL:2367656.2367672}, where a comparison between HLA and SOA concluded that:

\begin{itemize}
\item HLA has good interoperability, synchronization and effective and uniform information exchange mechanism between the communicating components (federates), but lacks several features of web services, such as: the integration of heterogeneous resources, web-wide accessibility across firewall boundaries;
\item SOA benefits from loose coupling, component reuse and scalability but lacks a uniform data exchange format and time synchronization mechanisms;
\item The combination of HLA and SOA can extend the capabilities of the two technologies and thus enable integrated simulated and real services.
\end{itemize}

Like HLA, the agent-directed simulation paradigm is used in a huge variety of fields but, when it comes to SimSaaS, few examples exists. Nevertheless, already in 2006, it was mentioned the importance of agents on simulation by exploring the relationship of software agents to simulation and games~\cite{Yilmaz01092006}. 

Some authors~\cite{Tolk:2010:UFA:2433508.2433550}\cite{Tolk:2011:MTI:2431518.2431551} defend that most of the current simulation interoperability standards are inadequate as they just support the federation by focusing on information exchange without providing the necessary introspective. HLA provides more flexibility as it only standardises how to structure the data, not the exchanging of information. However, the focus remains on the information exchanged within a system. Consequently, a formal approach to simulation interoperability using agent-supported simulation tries to solve this problem.

JAVA Agent DEvelopment Framework (JADE) is a common software framework that simplifies the implementation of multi-agent systems. Web Services Integration Gateway (WSIG) is an add-on for JADE which performs two-way translations between service requests and responses and JADE agent requests and responses. Thanks to this, it was possible to design a service-oriented simulation software framework as part of a broader approach towards generating improved levels of actionable views of situation awareness~\cite{Shao:2009:SSE:1639809.1639859}. 

Shao and McGraw referred that the great benefit of using JADE as the underlying agent development framework is that JADE agent entities can invoke web service functionality hosted outside the JADE run-time environment using normal JADE agent protocols, and that external entities can invoke JADE agent functionality from outside the JADE environment using normal web service protocols. Although the framework is very relevant, the applications were not in the cloud nor in a grid.

A truly implementation of agents in the cloud showed that agent-based M\&S can benefit a lot from cloud computing, making it easier to have more accurate and faster results, as well as timely experimentation and optimisation~\cite{Taylor:2014:TCC:2693848.2693884}. Even so, agent-based M\&S in the cloud may be highly complex due to the very different clouds, cloud middlewares and service approaches.

Federated simulation environments have some limitations in supporting dynamic model and simulating updating, as it was pointed in 2004 and 2006~\cite{Yilmaz01092006}. An example is HLA federation development, as it requires complete specification of object models and information exchanges before the simulation begins. It was also argued that there is a fundamental roadblock because of a lack of machine processable formal annotations describing behaviour, assumptions and obligations of federates.

\section{\uppercase{Taxonomy of the research work}}

SimSaaS is a trendy term which has been growing considerably in recent years. Thus, it is the ideal time to take advantage of this hype. However, there are some concerns: there is a lack of automation and integration of tools in M\&S~\cite{Wang:2015:SSM:2723553.2723554}, and research dissemination methods suffer as they do not allow publishing simulation code and scripts along with the published paper~\cite{Sliman2013611}. 

Herewith, HLA is another term referenced a lot in the literature since the first complete version (HLA 1.3) was published in 1998, but once again, when it comes to SimSaaS, almost nothing focuses on this and there is few work regarding extension of HLA to allow simulation services in general and in the cloud. Indeed, HLA solely has some disadvantages~\cite{Yilmaz01092006}\cite{Tolk:2010:UFA:2433508.2433550}.

In a 2014 panel about the future of research in M\&S~\cite{Yilmaz:2014:PFR:2693848.2694204} it is referred as a future research topic the distribution of SimSaaS in the cloud. So, once more SimSaaS is still mentioned as an unexplored area, now specifically in distributed simulation.

Although a truly implementation of agents in the cloud showed that agent-based M\&S can benefit from cloud computing~\cite{Taylor:2014:TCC:2693848.2693884}, there is a lack of work putting together agents and cloud in order to support SimSaaS. 

Summing up all the discoveries of the described literature review, it is possible to see a lot of gaps in the literature concerning SimSaaS, SimSaaS in specific domains of application, SimSaaS in the cloud, HLA in the cloud, solutions to HLA restrictions, agents to support SimSaaS and agents in the cloud. As the metaphor in the title of this paper tries to address, SimSaaS in the cloud is currently too sparse since it has so many gaps in research. It is necessary to make it less sparse (densifying) in order to augment the scientific and technological knowledge among researchers in the field. 

Densifying the sparse cloud SimSaaS is not just putting together cloud and SimSaaS, but also the synergies among them and agent-directed simulation and HLA, which could bring so many advantages. Fog Computing and Cloud 2.0, which were previously mentioned, could also help in this evolutionary process. Concluding, making these synergies a reality will be the front research opportunities for the next years.

A taxonomy of the research work could make the gaps identified and the research agenda for the next years more clear. Figure~\ref{fig:research_taxonomy} illustrates the taxonomy in the form of a Venn Diagram. Every work is about M\&S, more precisely SimSaaS. So, there are works that simply mention SimSaaS. Then, inside SimSaaS topic, research can focus on Cloud, HLA or Agent-directed simulation. In the particular case of Agent-directed simulation, there is a subset regarding Agent-supported simulation. As some works can address more than just one term, the representation in the form of a Venn Diagram was chosen to illustrate these possible synergies.

\begin{figure}[h!]
\centering
\resizebox{\linewidth}{!}{
\begin{tikzpicture}[fill=black!50]
	\draw (0,0) circle (1.5) (-1.5,0) node [text=black,left] {$Cloud$};
	\draw (1.5,0) circle (1.5) (3,0) node [text=black,right] {$AdSim$};
	\draw[green, thick, dashed] (1.5,0) circle (0.75) (1.3,-0.75) node [text=green,right,yshift=-4.5] {$AsSim$};
	\draw (0.75,1.5) circle (1.5)  (0.75,3) node [text=black,above] {$HLA$};
	\draw[specialblue, very thick] (-2.75,-2) rectangle (4.25,3.75) node [text=black,left,yshift=4.5] {$SimSaaS$};
	\draw[specialblue, very thick] (-3,-2.25) rectangle (4.5,4.5) node [text=black,left,yshift=4.5] {$M\&S$};
	\draw (0.5,-2.5) node [text=black,below] {\textit{(AdSim: Agent-directed Simulation, AsSim: Agent-supported Simulation)}};
\end{tikzpicture}
}
\caption{Diagram representing the taxonomy of the research work}
\label{fig:research_taxonomy}
\end{figure}

The presented taxonomy could be used as a conceptual framework for future developments. Researchers should look to these opportunities in the scientific community to orient their work, taking advantage of the benefits that this trends can give to their daily investigations.

\section{\uppercase{Conclusions}}

This paper started to briefly expose important concepts for a better understanding of its contents. After that, a literature review is presented, focusing in four distinct yet related topics: SimSaaS, Cloud Computing, HLA and Agents. Finally, a taxonomy of the research work is presented in the form of a Venn Diagram for a clear visualisation about which topics can (and should) have synergies among them. This taxonomy could be used as a conceptual framework for future developments.

With the front research opportunities for the next years and current research work identified, we hope this paper can leverage the scientific activity in the field, with researchers actually finding it useful to make the jump to the cloud. That jump will bring advantages not only to each researcher in particular, but also to the overall simulation scientific community seeking for more knowledge.

\section*{\uppercase{Acknowledgement}}
This work has been partially supported by MIEIC, Faculty of Engineering, University of Porto.

\bibliographystyle{apalike-modified}
{\small
\bibliography{literature-simultech}}

\end{document}